\documentstyle[amsmath,amssymb,graphicx]{article}

\def\be{\begin{eqnarray}}
\def\ee{\end{eqnarray}}
\def\nn{\nonumber}

\def\tr{{\rm tr}\,}

\hoffset= - 1.0in         % switch off for draft style
\title{{\bf Comment on integrability in Dijkgraaf-Vafa $\beta$-ensembles } \vspace{.2cm}}
\author{{\bf A.Mironov}\thanks{ {\small {\it
Lebedev Physics Institute} and {\it ITEP, Moscow, Russia}};
mironov@itep.ru; mironov@lpi.ru}, {\bf A.Morozov}\thanks{{\small
{\it ITEP, Moscow, Russia}}; morozov@itep.ru}, {\bf
Z.Zakirova}\thanks{{\small {\it Kazan Energy State University, Kazan, Russia}};
zolya\_zakirova@mail.ru}\date{ }}

\begin{document}
\maketitle

\vspace{-6.cm}

\begin{center}
\hfill FIAN/TD-01/12\\
\hfill ITEP/TH-07/12\\
\end{center}

\vspace{4.5cm}

\centerline{ABSTRACT}

\bigskip

{\footnotesize We briefly discuss the recent claims that the ordinary KP/Toda integrability,
which is a characteristic property of ordinary eigenvalue matrix models,
persists also for the Dijkgraaf-Vafa (DV) partition functions and
for the refined topological vertex.
We emphasize that in both cases what is meant is a particular representation
of partition functions: a peculiar sum over all DV phases in the first case
and hiding the deformation parameters in a sophisticated potential
in the second case, i.e. essentially a reformulation of some questions
in the new theory in the language of the old one.
It is at best obscure if this treatment can be made consistent with the
AGT relations and even with the quantization of the underlying integrable
systems in the Nekrasov-Shatashvili limit, which seem to require a full-scale
$\beta$-deformation of individual DV partition functions.
Thus, it is unclear if the story of integrability is indeed closed
by these recent considerations.}

\vspace{.5cm}

\section{Introduction}

Nowadays the abstract matrix model theory \cite{UFN1}
is once again on the rise.
One of the reasons for that is that the reformulation
of the Virasoro constraints or loop equations \cite{virco}
in terms of the AMM/EO topological recursion \cite{AMM/EO}
allowed to reveal hidden matrix model structures
in somewhat unexpected areas
like Seiberg-Witten theory and
conformal models (through the AGT relations \cite{AGT}) \cite{AGTmamo}
and knots \cite{knots}.
This poses the natural questions of how the other
properties of matrix models express themselves in these
circumstances.
The first in the line is, of course, integrability:
a mysterious fact that exact (non-perturbative)
partition functions in quantum field theory
satisfy {\it bilinear} relations
(while usual Ward identities, like Virasoro constraints,
provide only {\it linear} relations) \cite{qtau}.

The ordinary partition functions of eigenvalue matrix models
are typically the $\tau$-functions of the KP/Toda
type hierarchies \cite{UFN1,UFN2}.
Among other things, this fact is reflected in existence
of the Harer-Zagier recursion \cite{HZ},
a much more powerful than the ordinary AMM/EO one.
However, this property is lost
(or, better, modified in a still unknown way)
in the two important deviations:
after the $\beta$-deformation \cite{betadefo}
and in the Dijkgraaf-Vafa phases \cite{DVph}.
Recently there were claims to the opposite:
that integrable structure survives, moreover,
in both cases and presumably even in the combination of two.
The goal of this letter is to briefly comment on this
kind of statements.
We choose two particular examples: the papers
\cite{Sulk} on $\beta$-deformation
and \cite{BoEy} on the Dijkgraaf-Vafa phases.
In both cases the claim seems to reduce just to the
statement that deformed model can be considered
as a particular case of the non-deformed one,
thus, integrability of the ordinary Hermitian matrix model
implies bilinear relations for the deformed ones.
This is, of course, being a correct statement does not provide
any new interesting implications. In particular, this
does not help to construct any efficient Harer-Zagier recursion,
which would not be just a series in powers of $(\beta-1)$
or a result of peculiar summation over all the Dijkgraaf-Vafa phases.
We remind \cite{towaproof} that resolution of this problem
could provide a constructive interpretation
of the AGT relations as the Hubbard-Stratonovich duality \cite{HS}
in the doubly-quantized Seiberg-Witten theory (i.e. that in the $\Omega$-background
with the both non-zero deformation parameters\footnote{When only one $\epsilon$ is
non-vanishing, this corresponds to an ordinary quantization \cite{NSMM} of the underlying
integrable system \cite{SWint}.}).

\section{Integrability of Hermitian matrix model}

The old statement \cite{GMMMO,UFN1,UFN2} is that the integral
\be
Z_N = \frac{1}{N!} \prod_{i=1}^N \int d\mu_i  e^{V(\mu_i)} \Delta^2(\mu)
= \det_{ij} C_{i+j}
\label{detrep}
\ee
where Van-der-Monde determinant $\Delta(\mu) = \prod_{i<j} (\mu_i-\mu_j) = \det_{ij} \mu_i^{j-1}$
and the moment matrix
\be\label{mm}
C_i = \int d\mu\, e^{V(\mu)} \mu^{i-1}\equiv <\mu^{i-1}>
\ee
For $V(\mu) = V_0(\mu) + \sum_{k=0}^\infty t_k\mu^k$
one additionally has
\be
\frac{\partial C_i}{\partial t_j} = C_{i+j}
\ee
and the determinant representation (\ref{detrep}) along with this relation
is enough to demonstrate that $Z_N$ satisfies the Hirota bilinear
equations for the Toda chain $\tau$-function,
which, in turn, reduce to an infinite hierarchy of differential
equations, starting from\footnote{A more generic partition function \cite{GMMMO,UFN2}
$Z_N=\det C_{ij}$ depending on two sets of times $\{t\}$ and $\{\bar t\}$
is described by the two-dimensional Toda lattice hierarchy
with the first equation
$$
\frac{\partial^2 \log Z_N}{\partial t_1\partial \bar t_1} = {Z_{N+1}Z_{N-1}\over Z_N^2}
$$
provided $C_{ij}$ satisfies
$$
{\partial C_{ij}\over\partial t_k}=C_{i+k,j},\ \ \ \ \ \  \ \ \
{\partial C_{ij}\over\partial \bar t_k}=C_{i,j+k}
$$
In matrix models, it can be realized by an average $C_{ij}=<x^iy^j>$, which is the case
for multi-matrix models. Similarly, in the unitary matrix model case $C_{ij}=<x^{i-j}>$
and the partition function is a special reduction of the two-dimensional Toda lattice hierarchy
or, equivalently and even more naturally, of the two-component Toda hierarchy \cite{Unit}.
}
\be
\frac{\partial^2 \log Z_N}{\partial t_1^2} = {Z_{N+1}Z_{N-1}\over Z_N^2}
\label{Tc1}
\ee
Thus, one concludes that 
\be
Z_N=\tau\{N;t\}
\ee
What is important, these properties are independent of the choice
of the potential $V_0(\mu)$ and of the integration contours
in the definition of $C_i$ (one may say in a word that they do not depend on the
choice of measure).

Note that the $N!$ factor in the definition of $Z_N$ is essential: for the Gaussian
potential case, $V(\mu)=-{1\over 2g}\mu^2+t_1\mu$
\be
Z_N =\frac{1}{N!V_{U(N)}}\int_{N\times N} e^{-\tr V(M)} dM
=(2\pi)^{N/2} g^{N^2/2}\left(\prod_{k=1}^{N-1} k!\right)
\exp\left({\frac{gNt_1^2}{2}}\right)
\ee
and
\be
\frac{\partial^2 \log Z_N}{\partial t_1^2}=gN={Z_{N+1}Z_{N-1}\over Z_N^2}
\ee
The next Toda chain equation is the same as the first equation in KP hierarchy:
\be
3(\tau \tau_{22} - \tau_2^2)  - 4(\tau\tau_{13} - \tau_1\tau_3) +
(\tau\tau_{1111} - 4\tau_1\tau_{111} + 3\tau_{11}^2) = 0
\label{KP4}
\ee
where the index $i$ refers to the derivatives w.r.t. $t_i$.
One easily checks that the Gaussian Hermitian model satisfies this at the
point $\{{\rm all}\ t_k=0\}$ using formulas from \cite{CIVDV}. In these formulas
we preserve also the parameter $\beta$,
which would appear in the power of Van-der-Monde determinant in the eigenvalue
representation (\ref{detrep}), and put $V(\mu) = -\mu^2/2 + \sum_{k=0}^\infty t_k\mu^k$.
Then,
\be
\tau_1 = \tau_3 = \tau_{111} = 0, \nn \\
\tau_2 = \Big(\beta N^2 - (\beta-1)N\Big)\tau, \nn\\
\tau_{11} = N \tau, \nn \\
\tau_{22} = \Big(\beta^2N^4 - 2\beta(\beta-1)N^3 + (\beta^2+1)N^2 - 2(\beta-1)N\Big)\tau, \nn \\
\tau_{13} = 3\Big(\beta N^2 - (\beta-1)N\Big)\tau
\ee
Using these formulas one deduces that the l.h.s. of (\ref{KP4}) equals $-6(\beta-1)N(N-1)$
and vanishes when $\beta=1$.

\section{Sum over Dijkgraaf-Vafa phases}

The Dijkgraaf-Vafa phases emerge when the background potential $V_0(\mu)$
possesses several different extrema at points $\mu = \alpha_r$, $r=1,\ldots,s$.
Then the DV partition function is defined as a genus expansion around
the spectral curve, defined as a resolution of $y^2 = \left(V'_0(z)\right)^2 +  f(z)$
and depending on the $s$ extra moduli, hidden in the polynomial $f(z)$
of degree $(s-1)$.
As demonstrated in great detail in \cite{KMT,Mir,CIVDV} this definition is actually
equivalent to choosing $s$ different integration contours $K_r$,
so that $N_r$ out of $N$ eigenvalues $\mu_i$ are integrated along $K_r$.
These $N_r$ serve as the $s$ additional moduli, if the answer is analytically
continued from the integer values of $N_r$ to arbitrary ones.
Thus, one can define the Dijkgraaf-Vafa partition function
$Z_{N_1,\ldots,N_s}\{t_k\}$ as a matrix (or, better to say, eigenvalue)
model with $s$ different integration contours:
\be
Z_{N_1,\ldots,N_s}\{t_k\} = \prod_{r=1}^s \frac{1}{N_r!}
\left(\prod_{i=1}^{N_r}\int_{K_r} e^{V(\mu_i)} d\mu_i\right) \Delta^2(\mu)
\label{ZDV}
\ee
Now let us apply the determinant formula (\ref{detrep}) to this case:
\be
C_i = \sum_{r=1}^s e^{\xi_r} \int_{K_r} \mu^{i-1} e^{V(\mu)} d\mu
\ee
with arbitrary parameters $\xi_r$ (thus, the contour in (\ref{mm}) is given as a formal sum of
weighted contours, $\sum_{r=1}^s e^{\xi_r} K_r$).
Then, the Toda-chain tau-function is given by
\be
\tau(\vec\alpha,\vec\xi)\{t_k\} = \sum_{N_1,\ldots, N_s}
\left(\prod_{r=1}^se^{ N_r\xi_r}\right) Z_{N_1,\ldots,N_s}(\vec\alpha)\{t_k\}
\label{DVsummed}
\ee
One easily recognizes in this formula the sum (5.15) of ref.\cite{BoEy}.

To illustrate how the binomial coefficients are automatically taken into
account by the factorial in (\ref{ZDV}), we consider a very simple example of $N=2$.
For simplicity we also put $\xi_i=0$.
Then
\be
Z_2 = \frac{1}{2!} \prod_{i=1}^2 \left(\int_{K_1}+\int_{K_2}\right)
d\mu_i  e^{V(\mu_i)} \Delta^2(\mu)= \frac{1}{2!}\int_{K_1}
\int_{K_1}\prod_{i=1}^2 d\mu_i  e^{V(\mu_i)} \Delta^2(\mu)+\nn\\+
\frac{1}{1!1!}
\int_{K_1}\int_{K_2}\prod_{i=1}^2 d\mu_i  e^{V(\mu_i)} \Delta^2(\mu)
+ \frac{1}{2!} \int_{K_2}
\int_{K_2}\prod_{i=1}^2 d\mu_i  e^{V(\mu_i)} \Delta^2(\mu)=Z_{2,0}+Z_{1,1}+Z_{0,2}
\ee
or, in terms of the determinant representation (the indices of the averaging symbol $<...>$
enumerate contours):
\be
\det_{ij}C_{i+j}=\Big(<\mu^2>_1+<\mu^2>_2\Big)\Big(<1>_1+<1>_2\Big)-
\Big(<\mu>_1+<\mu>_2\Big)^2=\nonumber\\
={1\over 2!}\Big(2<\mu^2>_1<1>_1-2<\mu>_1^2\Big)+
\Big(<\mu^2>_1<1>_2+<\mu^2>_2<1>_2-2<\mu>_1<\mu>_2\Big)+
\nonumber\\
+{1\over 2!}\Big(2<\mu^2>_2<1>_2-2<\mu>_2^2\Big)=
 \frac{1}{2!}\int_{K_1}\int_{K_1}d\mu_1d\mu_2e^{V(\mu_1)}e^{V(\mu_2)}
\left(\mu_1-\mu_2\right)^2+\nonumber\\
+ \frac{1}{1!1!}\int_{K_1}\int_{K_2}d\mu_1d\mu_2e^{V(\mu_1)}e^{V(\mu_2)}
\left(\mu_1-\mu_2\right)^2
+ \frac{1}{2!}\int_{K_2}\int_{K_2}d\mu_1d\mu_2e^{V(\mu_1)}e^{V(\mu_2)}
\left(\mu_1-\mu_2\right)^2=
\ee

$$
=Z_{2,0}+Z_{1,1}+Z_{0,2}
$$

It is quite an exercise to check that (\ref{DVsummed}) made from
explicit expression for the DV partition function
$Z_{\vec N}(\vec \alpha)$ does indeed satisfy (\ref{Tc1}), (\ref{KP4})
and all other equations of the Toda chain and KP hierarchies,
in all orders of the genus expansion.
In fact, this check is somewhat similar to checking that theta-functions
satisfy these equations, by directly using their series expansions rather than
analytical properties, what is known to be a tedious exercise.
Still, some attempts of such direct checks were made in \cite{BoEy}.
We want to emphasize that the general argument,
that we just reminded in this section,
can be enough, provided one uses the demonstration of \cite{KMT,Mir,CIVDV}
that the Dijkgraaf-Vafa partition function $Z_{\vec N}(\vec \alpha)$ can be indeed
represented as a result of integrating some different eigenvalues
along different integration contours.

At the same time, despite (\ref{DVsummed}) is a $\tau$-function almost
trivially due to (\ref{detrep}),
this  sheds only some light on integrability
properties of DV partition functions $Z_{N_1,\ldots,N_s}$.
The claim is that these functions are a kind of Fourier
transform of the $\tau$-function, but in a very obscure
kind of variables: in $\vec\xi$, which from the point
of view of integrable hierarchies describe some
very non-explicit locus in the Universal Grassmannian
(the universal moduli space \cite{UMS}).
Even interpretation of these $\xi$'s in terms of Seiberg-Witten
theory remains obscure.
Thus this result calls for much better understanding before
it can be considered as a resolution of the problem
of integrability of DV partition functions.

\section{What would be the $\beta$-deformation of integrability theory?}

The recently discovered powerful AGT relation \cite{AGT} unifies \cite{AGTmore} the
three kinds of quantities,
which are {\it a priori} of a somewhat different origin:
the Nekrasov functions, conformal blocks
and peculiar $\beta$-ensembles of Dotsenko-Fateev or Penner type,
also known as "conformal matrix models".
There is little doubt that the basic underlying theory is
that of the $\beta$-Selberg integrals,
related to character expansion into the Jack and MacDonald polynomials
for $4d$ and $5d$ theories respectively.
A lot of these structures can actually be seen already at the
level of quantization of related integrable systems,
which is associated with the Nekrasov-Shatashvili
($\epsilon_2=0$) limit of the full $\Omega$-deformed
Seiberg-Witten (SW) structure.

However, there is an interesting option to treat the deformed
SW structure as an ordinary one.
This possibility is provided by the fact
that the ordinary SW equations
\be
\left\{\begin{array} {c}
a_i = \oint_{A_i}\Omega(z) \\
\frac{\partial F}{\partial a_i} = \oint_{B_i} \Omega(z)
\end{array} \right.
\ee
hold for the full prepotential $F(a|\epsilon_1,\epsilon_2)$
with $\epsilon_1,\epsilon_2\neq 0$ and any $\beta = -\epsilon_2/\epsilon_1$,
only with a sophisticated $\epsilon$-dependent SW differential
$\Omega(z|\epsilon_1,\epsilon_2)$
(which is actually a full, i.e. summed over all genera, 1-point
resolvent of the Dotsenko-Fateev $\beta$-ensemble,
to be provided by the yet unknown $\beta$-deformation of the Harer-Zagier recursion).

An intimately related observation \cite{Sulk} is that such a sophisticated representation
exists also for the
refined topological vertex \cite{rtv}, relevant to $\beta$-deformation of
Chern-Simons theory, and for the HOMFLY knot polynomials.

The problem is that such approaches hide all the relevant structures,
which one wants to {\it reveal} in the $\beta$-deformation,
in sophisticated quantities like $\Omega(z)$ or a sophisticated matrix model potential
$V(z)$,
and the problem is not resolved before the structure of these
quantities is fully understood.

Taking this to extreme, one may say that one can represent
the same quantity in two forms:
\be
\int \Delta^2(\lambda_i)\ \widetilde{d\mu}(\lambda_i)
= \int \Delta^{2\beta}(m_i)\ d\mu(m_i)
\label{ordbeta}
\ee
as ordinary matrix model and as a $\beta$-ensemble,
with a relatively simple measure $d\mu(m_i)$ and
a complicated measure $\widetilde{d\mu}(\lambda_i)$.
The two sides of this relation imply the two different ways to
switch on the time-variables:
insertion of $\prod_i \exp\sum_k\left( \tilde t_k \lambda_i^k\right)$
at the l.h.s. provides an ordinary $\tau$-function of the KP/Toda type,
while insertion of $\prod_i \exp\sum_k\left( t_k m_i^k\right)$
at the r.h.s. does not give rise to anything satisfactory.
The way to find an appropriate $\beta$-deformed version of this exponential
(and, more generally, of a $(\beta,q)$-exponential)
at the r.h.s. is exactly the problem of $\beta$-deformation of
integrability theory.

A very serious motivation for the study of relations like
(\ref{ordbeta}) is that the two measures at the two sides of the
equality are associated with different sets of symmetric functions:
the Schur and Jack polynomials respectively (and the MacDonald polynomials
would arise for the further $q$-deformed $\beta$-ensemble).
The point is that all these sets can be considered as different
bases in the space of symmetric functions and are therefore
linearly related (through the so called Kostka coefficients).
This means, first, that correlation functions in both representations
should indeed by somehow related, after the character expansion
technique is applied to express them through these polynomials.
Second, this makes the story of \cite{Sulk}
about the refined topological vertex
especially interesting, because in the associated theory
of superpolynomials \cite{superpols,DMMSS}
it is still unclear what is the preferred basis:
that of the MacDonald, of Schur or rather of the Hall-Littlewood polynomials \cite{MMSS}.
Amusingly a place is still not found there for the Jack polynomials,
putting under a big question the literal $\beta$-ensemble approach
to topological vertices (this is also illustrated by the failure
to generalize the Chern-Simons matrix model for torus HOMFLY
polynomials to the case of superpolynomials by switching
from ordinary matrix models to $\beta$-ensembles \cite{GalMirMor}).

\section{Conclusion}
To conclude, we tried to argue that integrability properties of
$\beta$-deformed and weighted-averaged Dijkgraaf-Vafa partition functions
remain an important and deep problem, which is still far from
being solved.
The results of \cite{Sulk} and \cite{BoEy} respectively,
in this direction are very important, but seem to reflect only
straightforward consequences of those of Hermitian matrix model.
In this sense they do not provide any new non-trivial information
about above deformations.
In particular, they do not yet help to generalize the Harer-Zagier recursion
and matrix model interpretation of the AGT relations
and knot polynomials to the case of $\beta\neq 1$.
However, it also remains an open question,
if any less trivial deformations
of integrability properties exist at all in these cases.

\section*{Acknowledgements}

Our work is partly supported by Ministry of Education and Science of
the Russian Federation under contract 02.74.11.0081, by NSh-3349.2012.2,
by RFBR grants 10-02-00509 (A.Mir. and Z.Z.), 10-02-00499 (A.Mor.) and
by joint grants 11-02-90453-Ukr, 12-02-91000-ANF,
12-02-92108-Yaf-a, 11-01-92612-Royal Society.

\end{document}